\documentclass[apjl]{emulateapj}
\usepackage{apjfonts}
\usepackage{natbib}

\slugcomment{Accepted for publication in ApJ}

\shorttitle{Gamma-ray Emission from SNR S147}
\shortauthors{J. Katsua et al.}

\begin{document}

\title{\emph{Fermi}-LAT Observation of Supernova Remnant S147}

\author{
J.~Katsuta\altaffilmark{2,1}, 
Y.~Uchiyama\altaffilmark{2,1}
T.~Tanaka\altaffilmark{2}, 
H.~Tajima\altaffilmark{2,7}, 
K.~Bechtol\altaffilmark{2}, 
S.~Funk\altaffilmark{2}, 
J.~Lande\altaffilmark{2}, 
J.~Ballet\altaffilmark{3}, 
Y.~Hanabata\altaffilmark{4}, 
M.~Lemoine-Goumard\altaffilmark{5,6}, 
T.~Takahashi\altaffilmark{8}
}
\altaffiltext{1}{Corresponding author: J.~Katsuta, katsuta@slac.stanford.edu; Y.~Uchiyama, uchiyama@slac.stanford.edu.}
\altaffiltext{2}{W. W. Hansen Experimental Physics Laboratory, Kavli Institute for Particle Astrophysics and Cosmology, Department of Physics and SLAC National Accelerator Laboratory, Stanford University, Stanford, CA 94305, USA}
\altaffiltext{3}{Laboratoire AIM, CEA-IRFU/CNRS/Universit\'e Paris Diderot, Service d'Astrophysique, CEA Saclay, 91191 Gif sur Yvette, France}
\altaffiltext{4}{Department of Physical Sciences, Hiroshima University, Higashi-Hiroshima, Hiroshima 739-8526, Japan}
\altaffiltext{5}{Universit\'e Bordeaux 1, CNRS/IN2p3, Centre d'\'Etudes Nucl\'eaires de Bordeaux Gradignan, 33175 Gradignan, France}
\altaffiltext{6}{Funded by contract ERC-StG-259391 from the European Community}
\altaffiltext{7}{Solar-Terrestrial Environment Laboratory, Nagoya University, Nagoya 464-8601, Japan}
\altaffiltext{8}{Institute of Space and Astronautical Science, JAXA, 3-1-1 Yoshinodai, Chuo-ku, Sagamihara, Kanagawa 252-5210, Japan}

\begin{abstract}
We present an analysis of gamma-ray data obtained with 
the Large Area Telescope (LAT) onboard the \emph{Fermi Gamma-ray Space Telescope} in the region around SNR S147 (G180.0$-$1.7). 
A spatially extended gamma-ray source detected in an energy range of 0.2--10~GeV is found to coincide with SNR S147.
We confirm its spatial extension at $>$$5\sigma$ confidence level.
The gamma-ray flux is $(3.8 \pm 0.6) \times 10^{-8}$~photons~cm$^{-2}$~s$^{-1}$, corresponding to 
a luminosity of $1.3 \times 10^{34}$~($d/1.3$~kpc)$^2$~erg~s$^{-1}$ in this energy range. 
The gamma-ray emission exhibits a possible spatial correlation with prominent H$\alpha$ filaments of S147.
There is no indication that the gamma-ray emission comes from the associated pulsar PSR J0538$+$2817.
The gamma-ray spectrum integrated over the remnant is likely dominated by the decay of neutral $\pi$ mesons produced through the proton--proton collisions in the filaments. 
Reacceleration of pre-existing CRs and subsequent adiabatic compression in the filaments is sufficient to provide the required energy density of high-energy protons.
\end{abstract}


\section{Introduction}
\label{sec: Intro}

GeV gamma-ray sources associated with middle-aged supernova remnants (SNRs)
interacting with molecular clouds
have recently been discovered with the Large Area Telescope (LAT) onboard the \textit{Fermi Gamma-ray Space Telescope}~\citep{FermiW51C,FermiW44,FermiIC443,FermiW28,Slane}. 
The GeV emission of these SNRs is plausibly dominated by the decay of $\pi^0$ mesons created by proton-proton collisions, 
although a dominant electron bremsstrahlung component is an alternative interpretation~\citep{FermiW51C}. 
The SNRs interacting with molecular clouds are found to 
be more luminous gamma-ray sources ($\sim10^{35\mbox{--}36}\ \rm erg\ s^{-1}$ in the LAT band) than other types of SNRs \citep[see][]{Thompson11}.
The observed luminosity requires high-density gas so that collisions between relativistic particles and target gas become efficient enough, which is 
easily explained by interactions with molecular clouds.
Two different types of scenarios are put forward to explain the bright gamma-ray emission. 
One scenario explains the gamma rays as the $\pi^0$-decay emission due to interactions between nearby molecular clouds and relativistic protons escaping from the SNR system~\citep{AA96}. 
Another scenario considers the $\pi^0$-decay emission from shock-compressed clouds in a middle-aged SNR,
where the accelerated particles frozen in the shocked clouds efficiently collide with target gas~\citep{Uchiyama10}. 

Recently, GeV gamma-ray emission from the Cygnus Loop, a middle-aged 
remnant without a clear signature of interactions with molecular clouds, 
has been detected with the \emph{Fermi} LAT \citep{FermiCygLoop}. 
The gamma-ray luminosity is quite low ($\sim10^{33}\ \rm erg\ s^{-1}$) compared to the SNRs with molecular cloud interactions. 
The gamma-ray emission can be interpreted to come from diffuse gas behind the blast wave, but dense optical filaments seen in an H$\alpha$ image 
would also offer a plausible site of the gamma-ray production.
Detections of other low-luminosity SNRs with the \emph{Fermi} LAT, 
thanks to an ever increasing exposure time, will help understand 
the gamma-ray production sites and constrain cosmic-ray (CR) acceleration in such SNRs.

SNR S147 (G180.0$-$1.7), located toward the Galactic anticenter, is one of the most evolved SNRs in our Galaxy. 
S147 has a nearly circular shape with an angular diameter of $\simeq 200\arcmin$. 
It contains a complex network of long filaments in the optical band~\citep{S147_k1,S147_Ha}, which are also visible in the radio band~\citep{S147_R2,S147_R3,S147_R1}.
These observations also show that the radio and H$\alpha$ emissions 
correlate very well.
The velocity of the optical filaments is estimated to be 80--120~km~s$^{-1}$ 
\citep{S147_k1,S147_k2,S147_k3}.
No indication of interactions with molecular clouds has been reported.

The radio flux density is 70~Jy at 1~GHz, and the radio spectrum is known to have a spectral break around 1.5~GHz.
\cite{S147_R1} obtained the integrated spectrum of S147 in the 0.1--5~GHz range, and determined the spectral indices $\alpha \sim0.30\pm0.15$ and $\alpha \sim1.20\pm0.3$
below and above the spectral break, respectively.
\cite{S147_R1} also found that the filamentary and diffuse emission in S147 have different spectral indices of $\alpha \sim0.35$ and $\sim1.35$ respectively in the frequencies of 2.6--4.8~GHz (above the break).
The radio emission is considered to be synchrotron radiation from high-energy electrons, while the origin of the spectral break is uncertain.
No X-ray emission has been reported to date from this region~\citep{S147_X} nor any TeV emission.
The EGRET detected a GeV gamma-ray source (3EG J0542+2610) in the vicinity of S147, 
but its association with the SNR was unclear due to its large positional uncertainty (0\fdg 70 at 95\% confidence level)~\citep{EGRET3rd}.

PSR~J0538$+$2817 plausibly associated with S147 is located near the center of the SNR~\citep{S147_psr_parallax}. 
The pulsar has a spin period of 143~ms and a spin-down luminosity of $5\times10^{34}$~erg~s$^{-1}$. 
It was first discovered in the radio~\citep{S147_psr_R} and was later also found in the X-rays~\citep{S147_psr_X2}.
The X-ray observation also revealed extended emission indicative of a pulsar wind nebula (PWN)~\citep{S147_psr_X1}.
Estimated pulsar ages differ significantly between a kinematic age of $\sim3 \times 10^4$~yr~\citep{S147_psr_pm} and a characteristic age of $6\times10^5$~yr~\citep{S147_psr_R}.
The kinematic age of the pulsar is broadly consistent with estimates of the SNR age which are in the range of (2--10)$ \times 10^4$~yr~\citep{S147_R3,S147_R2}.

The distance to the SNR is likely to be $d=1.3$~kpc given a plausible association with pulsar PSR~J0538$+$2817.
The distance to the pulsar is estimated to be $1.30^{+0.22}_{-0.16}$~kpc from the parallax~\citep{Chatterjee09} and $1.2\pm0.2$~kpc with a dispersion measure~\citep{S147_psr_DM}.
On the contrary, $d<0.88$~kpc is suggested by absorption lines of the B1e star HD36665 which originates in gas associated with S147~\citep{S147_k3,S147_star}.
The diameter of the SNR is $D \simeq 76\ (d/1.3\ {\rm kpc})$~pc.

Here we report the LAT observations of SNR S147. 
A GeV gamma-ray source spatially coincident with S147 is designated as 1FGL~J0538.6+2717 in the \emph{Fermi} LAT First Source Catalog (1FGL catalog)
published by the \emph{Fermi} LAT Collaboration~\citep{1FGL}.
In this paper, we present detailed analysis of this LAT source with much longer accumulation time of about 31 months. 
This paper is organized as follows. In section 2, the observations with the \emph{Fermi} LAT and the data reduction are summarized. 
Analysis of the LAT source in the direction of SNR~S147 
is reported in section 3, establishing an association between the gamma-ray source and SNR S147. 
In section 4, we present modeling of the gamma-ray emission coming from SNR S147 and discuss its implications to the CR acceleration in middle-aged SNRs.

\section{Observation and Data Reduction}

The \emph{Fermi Gamma-ray Space Telescope} was launched on 2008 June 11. 
The LAT onboard \emph{Fermi} is a pair conversion telescope, equipped with solid state silicon trackers and cesium iodide calorimeters, sensitive to photons in a broad energy band from 0.02 to $>$$300$~GeV. The LAT has a large effective area ($\sim$8000 cm$^2$ above 1 GeV for on-axis events), instantaneously viewing $\sim$2.4 sr of the sky with a good angular resolution (68\% containment radius better than $\sim$1$^{\circ}$ above 1~GeV). 
Details of the LAT instrument and data reduction are described in \citet{LAT}.

The LAT data used here were collected for about 31 months from 2008 August 4 to 2011 March 1. 
The \emph{Diffuse} event class was chosen
and photons beyond the earth zenith angle of 105$^\circ$ were excluded to reduce the background from the Earth limb.
The {\sf P6\_V11}\footnote[1]{http://www.slac.stanford.edu/exp/glast/groups/canda/archive/ pass6v11/lat\_Performance.htm} instrument response functions were used for the analyses in this paper.

\section{Analysis and Results}

We utilized {\sf gtlike} in the Science Tools analysis package\footnote[2]{available at the \emph{Fermi} Science Support Center. http://fermi.gsfc.nasa.gov/ssc} for spectral fits.
With {\sf gtlike}, a binned maximum likelihood fit is performed 
on the spatial and spectral distributions of observed gamma rays to optimize spectral parameters of the input model taking into account the energy dependence of the point-spread function (PSF).

\begin{figure*}[h] 
\begin{center}
\includegraphics[width=7.5in]{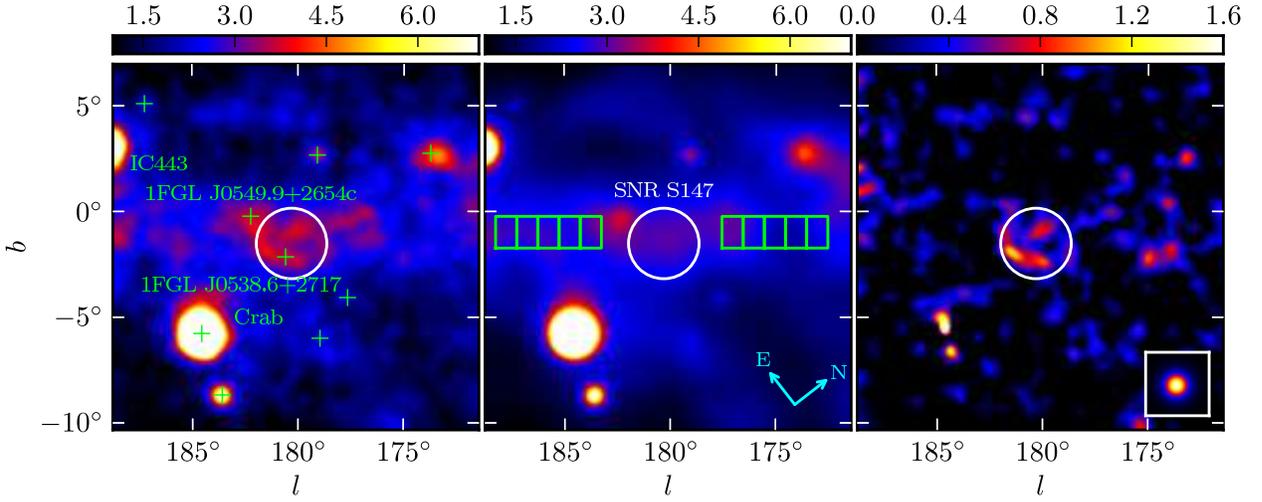}
\caption{
(Left) 
\textit{Fermi}-LAT count map above 1 GeV around SNR S147 in units of counts per pixel. The pixel size is $0\fdg 1$.
Smoothing with a  Gaussian kernel of $\sigma =0 \fdg 25$ is applied. 
SNR S147 is represented by a white circle. 
The background sources contained in the ROI are shown as green crosses. 
(Middle)
Background model map.
The green boxes each with the dimensions of $1 \fdg 0 \times 1 \fdg 5$ represent the regions used for the evaluation of the accuracy of the Galactic diffuse model. 
(Right)
Background-subtracted count map. A simulated point source is shown in the inset.
\label{fig: largeFoV}}
\end{center}
\end{figure*}

Analysis using {\sf gtlike} is performed in a $17^\circ\times17^\circ$ region around SNR S147, 
which is referred to as a region of interest (ROI). 
The fitting model includes other point sources whose positions are given in the 1FGL catalog.\footnote[3]{The flux of S147 would change $<10$\% below 1~GeV and $<4$\% above 1~GeV when we use 2FGL catalog~\citep{2FGL}.}
Since the nearby SNR IC443 is spatially resolved by the LAT,
we model its spatial distribution as a disk centered at ($\alpha$, $\delta$)=($94\fdg 31$, $22\fdg 58$) with a diameter of $48\arcmin$ according to \cite{FermiIC443}.
The Galactic diffuse emission is modeled by ``gll\_iem\_v02\_P6\_V11\_DIFFUSE.fit" and an isotropic component (instrumental and extragalactic diffuse backgrounds) by ``isotropic\_iem\_v02\_P6\_V11\_DIFFUSE.txt".
Both background models are the standard diffuse emission models released by the LAT team\footnotemark[2].
In each {\sf gtlike} run, all point sources within the ROI and diffuse components in the model are fitted with the normalization left free.
The spectral shapes are either fixed or set free depending upon the specific analysis (see below).
Note that we also include 1FGL sources outside the ROI but within $14^\circ$ from S147 with their parameters being fixed at those of the 1FGL catalog.
The count map of the ROI above 1~GeV (Figure~\ref{fig: largeFoV} left) shows that the Crab Nebula and SNR~IC443 are the dominant gamma-ray sources in this region. 
We model the Crab spectrum using three components 
(the Crab pulsar, the Crab PWN synchrotron, and the Crab PWN inverse Compton components), and fix their spectral shapes in the fit, following the previous LAT study~\citep{FermiCrab}.
For IC443, we model the emission as a broken power law following~\cite{FermiIC443}, and also fix its spectral shape.

\begin{figure*}[H] 
\begin{center}
\includegraphics[width=5in]{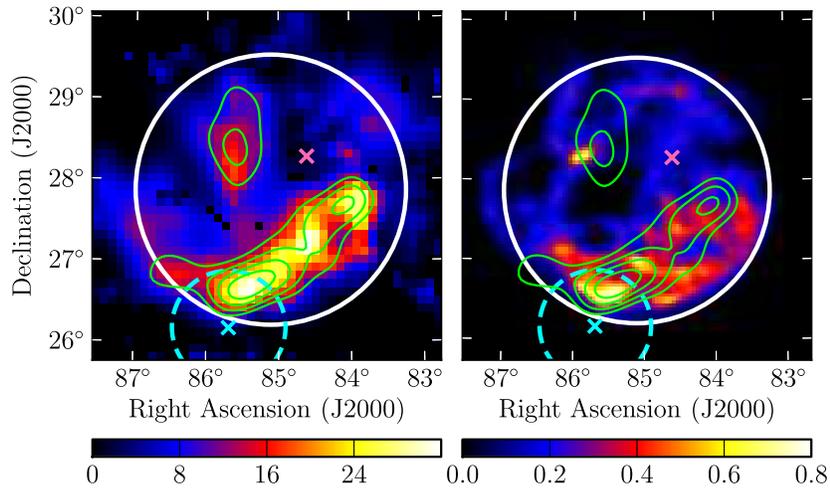}
\caption{
Left: TS map obtained with maximum likelihood analysis of the \emph{Fermi}-LAT data in the vicinity of SNR S147 above 1~GeV. 
Overlaid are linear contours of the background-subtracted count map 
above 1~GeV taken from Figure~\ref{fig: largeFoV}. 
A white circle  represents the outer boundary of SNR S147. 
A cyan cross and circle represent a position and positional error (95\% confidence level) of 3EG J0542+2610, respectively~\citep{EGRET3rd}.
A magenta cross indicates the position of PSR J0538+2817. 
Right: H$\alpha$ flux intensity map of SNR~S147 in units of rayleighs $(10^6/4\pi)$~photons~cm$^{-2}$~s$^{-1}$~sr$^{-1}$~\citep{S147_Ha_map}, 
with the contours of the background-subtracted count map overlaid.
\label{fig: smallFoV}}
\end{center}
\end{figure*}

\subsection{Spatial Distribution}
\label{sec: detection}

\begin{figure*}[t] 
\begin{center}
\includegraphics[width=5in]{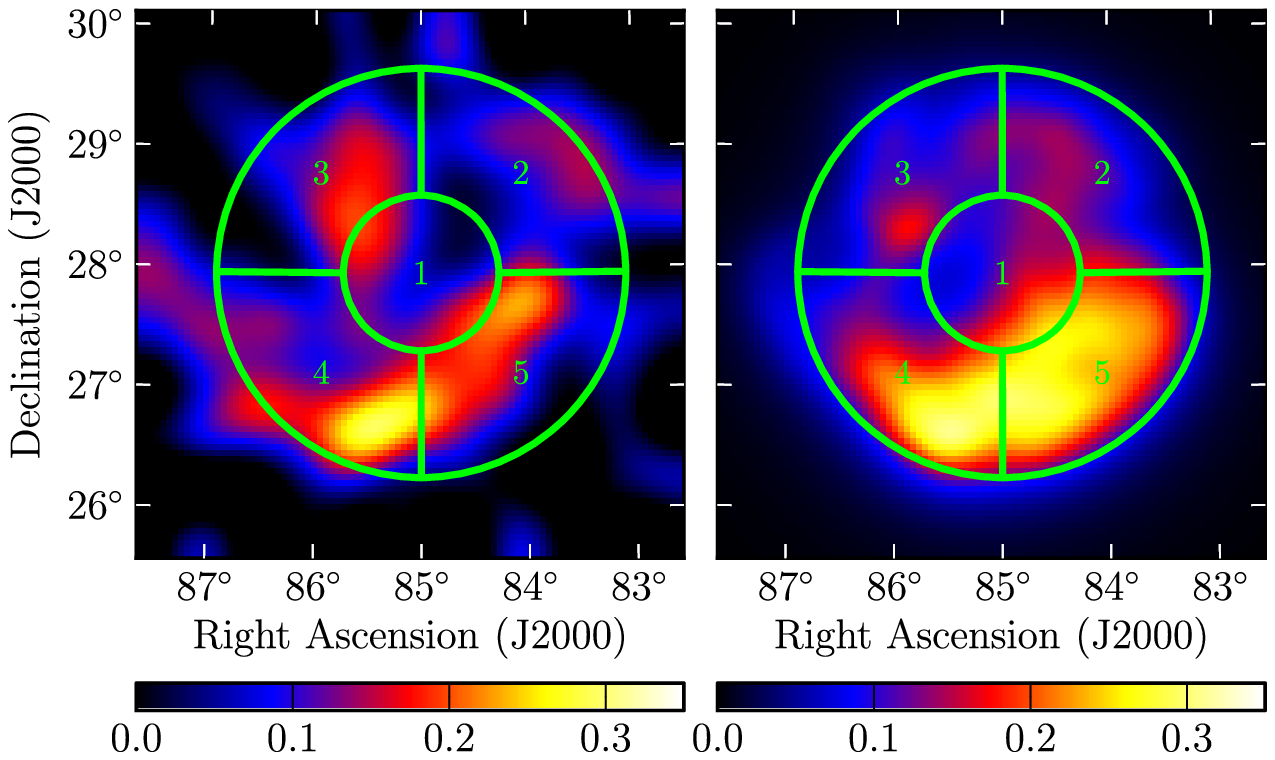}
\caption{\small
Five divided regions for comparison of gamma-ray and H$\alpha$ emission. 
The grids overlay (Left) the LAT count map in the vicinity of SNR~S147 above 1~GeV with the background subtracted, 
and (Right) an H$\alpha$ image convolved with the LAT PSF (see Section~\ref{sec: detection}).
The count map and image is in units of counts per pixel with the size of $0\fdg 05$.
Here the LAT count map is smoothed by a Gaussian kernel of $\sigma =0 \fdg 25$, 
to show the clear spatial distribution of the S147 source with the low number of photon counts.
Note that the the count map is not smoothed for the comparison of each region.
\label{fig: grid}}
\end{center}
\end{figure*}

In Figure~\ref{fig: largeFoV}, we show a smoothed count map of the ROI above 1~GeV, 
a corresponding background model map, and the background-subtracted count map.
The background model map includes contributions from the Galactic diffuse emission, isotropic diffuse background, and nearby discrete sources.
The model parameters of diffuse components and nearby sources are set at the best-fit values obtained by {\sf gtlike} using the data above 1~GeV.
An extended gamma-ray source associated with S147 is visible in the background-subtracted map.
In the model, the spatial distribution of the S147 source is assumed to be the H$\alpha$ image~\citep{S147_Ha_map} (see below). 
The spectral shape is fitted as a power-law function with the index set free.
The point source in the inset is simulated with the same spectral shape as S147 obtained by maximum likelihood fits.

We generated a Test Statistic (TS) map using the LAT data above 1 GeV in the S147 region (Figure~\ref{fig: smallFoV} left).
TS is defined as $-2\Delta\ln ({\rm likelihood})$ obtained by {\sf gtlike} between models of the null hypothesis and an alternative.
In this paper, we refer to TS as a comparison between models without a target source (null hypothesis) and with the source (alternative hypothesis) unless otherwise mentioned.
In this map, the TS value at each grid position is calculated by using a model with a point source placed at the position, which has a power-law energy distribution with its index being free.
The excess gamma rays above backgrounds are distributed inside the SNR boundary and the spatial extent is consistent with the remnant size.
To evaluate the spatial extent, we fit the data with a model where the spatial distribution of the S147 source is set to a disk image with a uniform surface brightness, the size of which corresponds to that of the SNR in the H$\alpha$ image (the white circle in Figure~\ref{fig: smallFoV}).
Indeed, TS obtained with this model is larger by $\sim50$ than the largest TS of a point source in the S147 region above 1~GeV,
which means the S147 source is extended with $>$$5\sigma$ confidence. 
The spectral distribution of the S147 source is assumed to be a power-law function with its index free.

In Figure~\ref{fig: smallFoV} (right), 
the background-subtracted gamma-ray map is compared with an H$\alpha$ emission map of S147. 
To perform a more detailed comparison between the gamma-ray and H$\alpha$ maps, 
we divide the S147 region into five cells as shown in Figure~\ref{fig: grid}, and 
compare gamma-ray ($>$1~GeV) and H$\alpha$ fluxes in each cell. 
We note that the size of the cell ($\sim 1^\circ$) is comparable to the LAT angular resolution (better than $\sim$1$^{\circ}$ above 1~GeV).
To take into account the LAT PSF for the morphological comparison,
the H$\alpha$ image is convolved with the LAT PSF as shown in Figure~\ref{fig: grid}.
The gamma-ray map used for this comparison is a count map with all model components except for the S147 source subtracted as described above. 
Since the PSF depends on energy, we calculate the PSF assuming the spectral index of the S147 source to be 2.5, 
the best-fit value obtained with the H$\alpha$ image used as a template above 1~GeV.
Note that the exposure of the LAT observations is uniform within 0.5\% in this region.

\begin{figure}[h] 
\includegraphics[width=3.5in]{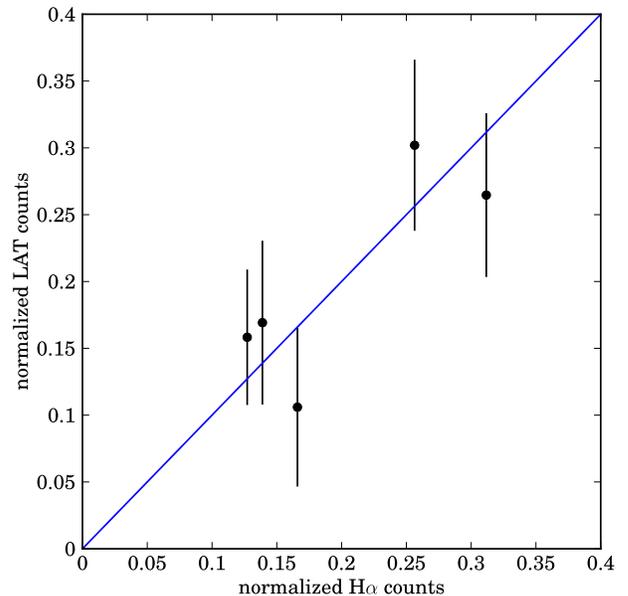}
\caption{\small
Relation of the LAT (1--200~GeV) and H$\alpha$ counts. Note that both are normalized to make their total counts 1.0.
The blue line represents a linear function with its slope of 1.0.
The error bars represent statistical uncertainty associated with the LAT-observed photon counts ($1\sigma$).
\label{fig: cor}}
\end{figure}

Figure~\ref{fig: cor} shows the resulting correlation diagram. 
The plotted gamma-ray counts are obtained by summing up the subtracted count map of the LAT in each region,
while the H$\alpha$ counts are calculated by summing up the H$\alpha$ image convolved with the LAT PSF.
The figure shows a possible correlation between gamma-ray and H$\alpha$ fluxes.
Though it is challenging to establish the presence of the correlation given large statistical uncertainties, 
the diagram suggests that an extension of the {\it Fermi} mission ($\sim10$~yr) will provide an opportunity to confirm it.

\begin{figure*}
\begin{center}
    \includegraphics[width=5in]{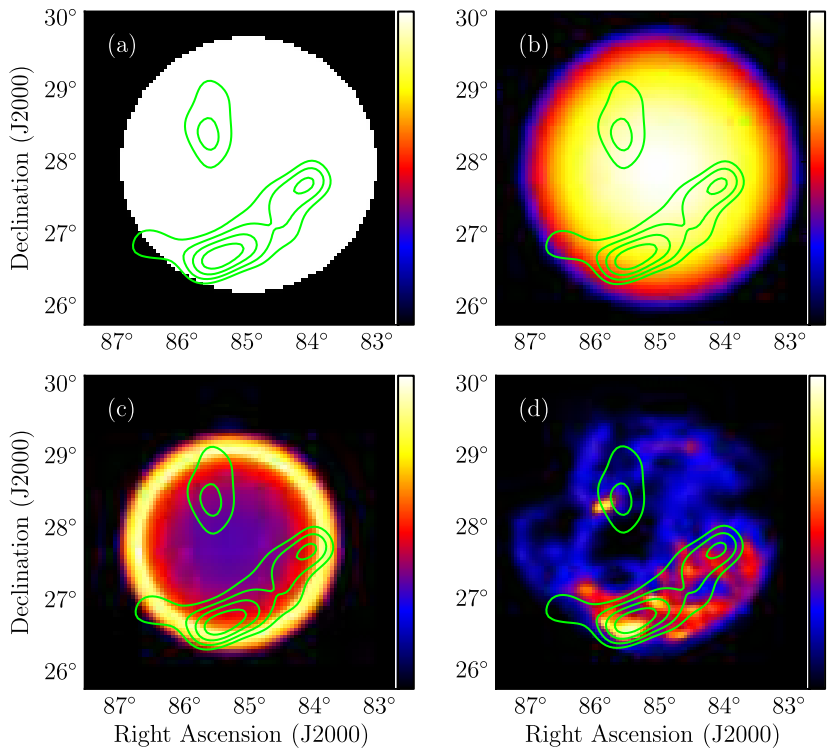}
  \caption{\small The S147-source templates for the LAT spectral analysis. 
  The templates are (a) a disk template, (b) a sphere template, (c) a shell template, and (d) an H$\alpha$ image, overlaid with 
  the contours of the LAT background-subtracted count map above 1~GeV (the same contours as that overlaid in Figure~\ref{fig: smallFoV}).}
  \label{fig:S147:templates}
  \end{center}
\end{figure*}

We also test different spatial templates, a disk template, sphere template, and shell template, as shown in Figure~\ref{fig:S147:templates}.
The disk template is a disk with uniform surface brightness.
The sphere template is a two-dimensional projection of a sphere with a uniform emissivity per unit volume.
The shell is a two-dimensional projection of a spherical shell with a ratio of inner diameter to outer diameter set to 0.8 based on the H$\alpha$ map.
Center positions and diameters of the disk, sphere, and shell templates are fitted to the data.
Using the different templates, we perform maximum likelihood fits and compare the best-fit parameters in the energy range of 1--200~GeV.
The spectral shape of the S147 source in the wide energy range is assumed to be a power-law function.
As shown in Table~\ref{tbl: TSs}, the S147 source is significantly detected in each case, and the obtained fluxes and spectral shapes are almost the same.
The H$\alpha$ image has the largest TS among all templates, despite the fact that the other templates have three more free parameters (position and diameter) than those of the H$\alpha$ image.

\begin{deluxetable*}{lccccc}
\tabletypesize{\scriptsize}
\tabletypesize{\footnotesize}
\tablecaption{Best-fit Values of Different Templates for the S147 Source\label{tbl: TSs}}
\tablewidth{0pt}
\tablehead{
\colhead{Template} & \colhead{Center position} & \colhead{Diameter} & \colhead{Flux} & \colhead{Photon index} & \colhead{TS}\\
\colhead{} & \colhead{($\alpha$, $\delta$)} & \colhead{} & \colhead{[1$0^{-9}$~photons~cm$^{-2}$~s$^{-1}$]} & \colhead{} & \colhead{}
}
\startdata
Disk       &  ($84\fdg 97$, $27\fdg 97$) &  $3\fdg 5$  & $9.2\pm1.1$ & $2.45\pm0.14$ & 86.6\\
Sphere  &  ($85\fdg 00$, $27\fdg 93$) &  $3\fdg 9$  & $9.2\pm1.1$ & $2.45\pm0.14$ & 84.7\\
Shell      &  ($85\fdg 29$, $27\fdg 80$) &  $3\fdg 1$  & $8.6\pm1.0$& $2.51\pm0.15$&  88.0\\
H$\alpha$ image & -- & -- & $8.5\pm1.0$ & $2.48\pm0.14$& 94.8\\
\enddata
\tablecomments{Flux of the S147 source is calculated in the energy range of 1--200~GeV.}
\end{deluxetable*}

Given the results of the correlation diagram and the fits of the different spatial templates, 
the gamma-ray emission from the S147 source has a possible spatial correlation with the H$\alpha$ filaments.
Hereafter we adopt the H$\alpha$ image as the spatial template of the S147 source.

\subsection{Spectrum}
\label{sec: sed}

The spectral energy distribution (SED) of the source associated with S147 is obtained by dividing the 0.2--200~GeV energy band into six energy bins. 
Since each energy range for the fit is small, 
we model the Crab as a point source with a single spectral component instead of three.
To model the bright emissions from Crab and IC443 in the small energy range,
these spectral shapes are adopted to be power-law functions with their indices free.
The indices of the other sources are fixed at the values in the 1FGL catalog. 
S147 is fitted with a simple power-law function in each energy bin with its photon index fixed at 2.1, which is obtained by a broadband fitting in 0.2--200~GeV (see below).
We note that the flux obtained for S147 in each energy bin is insensitive to the choice of index, if it is confined to a reasonable range (say, 1--3). 
In each fit, all sources within the ROI and diffuse components in the model are fitted with their normalization being free.

The systematic errors in the spectral analysis are mainly due to uncertainties associated with the underlying Galactic diffuse emission and uncertainties of the effective area of the LAT.
The uncertainties of the Galactic diffuse emission are primarily due to the imperfection of the Galactic diffuse model and/or the contributions from discrete sources not resolved from background.
We evaluate the uncertainties of the Galactic diffuse emission
by measuring the dispersion of the fractional residuals in 10 regions around S147, where  the Galactic diffuse component dominates (Figure~\ref{fig: largeFoV}).
The fractional residuals, namely (observed$-$model)/model, are calculated in three energy bands (0.20--0.45~GeV, 0.45--1.0~GeV, and 1.0--10~GeV) for each region. 
At each energy range, the uncertainties of the Galactic diffuse model are adopted separately as the second largest values among 10 residuals (90\% containment).
From the results, the uncertainties are evaluated as 5.3\%, 6.4\%, and 8.9\% at energy ranges of 0.20--0.45~GeV, 0.45--1.0~GeV, and 1.0--10 GeV, respectively.
Systematic uncertainties of the effective area are 10\% at 100 MeV, decreasing to 5\% at 560 MeV, and increasing to 20\% at 10 GeV and above~\citep{LATcal}.
Note that systematic errors associated with choice of spatial models of S147 are negligible, 
since the best-fit parameters obtained by {\sf gtlike} with the different spatial templates for S147 are almost the same as described in Section~\ref{sec: detection}.

\begin{figure}[h] 
\includegraphics[width=3.5in]{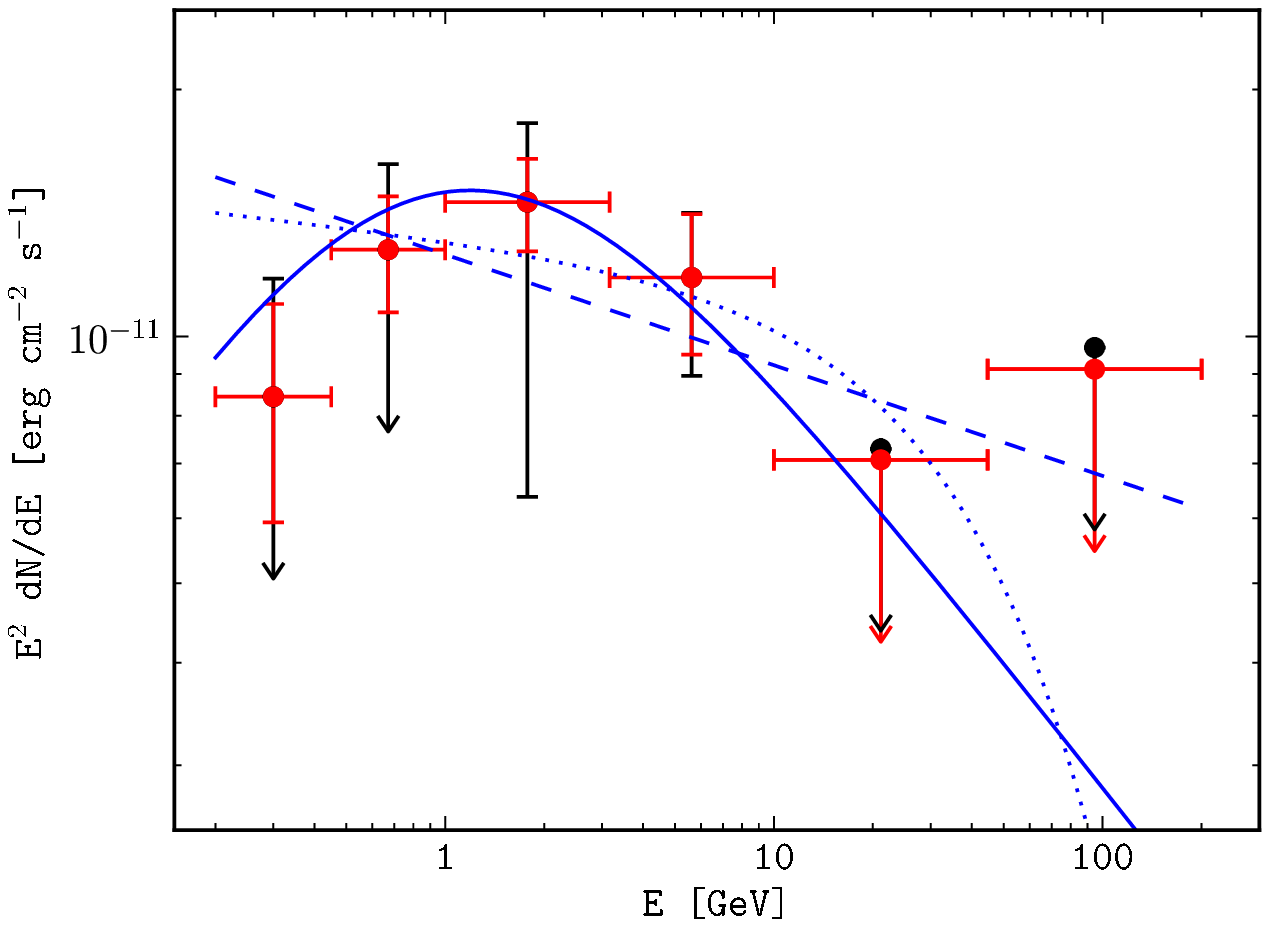}
\caption{\small
SED of S147 measured by the \textit{Fermi} LAT. 
The arrows represent upper limit on the fluxes at 90\% confidence level.
The total systematic errors are indicated by black error bars while statistical errors (1$\sigma$) are indicated by red error bars.
The black points and arrows represent upper limits taking the systematic errors into consideration.
The blue dashed, dotted, solid lines represent the best-fit power-law, exponentially cut-off power-law, smoothly broken power-law functions from a binned likelihood fit in 0.2--200 GeV.
\label{fig: sed}}
\end{figure}

Figure~\ref{fig: sed} shows the resulting SED for S147.
Total systematic errors are set by adding in quadrature the uncertainties due to the Galactic diffuse model and the effective area.
Note that upper limits are calculated for the spectrum of S147 below 1~GeV given the conservative systematic errors.
Inspection of the figure indicates a spectrum steepening above a few GeV.
To evaluate the significance of the steepening, we performed likelihood-ratio tests between a power-law function (the null hypothesis), and 
either an exponentially cut-off power-law or a smoothly broken power-law function (the alternative hypotheses) for 0.2--200~GeV data.
The exponentially cut-off power-law function is described as
\begin{equation}
\frac{dN}{dE}=N_0\left(\frac{E}{E_0}\right)^{-\Gamma} \exp{\left(-\frac{E}{E_{\rm cut}} \right)},
\label{eq:g-cutoff}
\end{equation}
where $E_0$ is 1~GeV. The photon index $\Gamma$, a cutoff energy $E_{\rm cut}$, and a normalization factor $N_0$ are free parameters.
The smoothly broken power-law function is described as
\begin{equation}
\frac{dN}{dE}=N_0\left(\frac{E}{E_0}\right)^{-\Gamma_1}\left(1+\left(\frac{E}{E_{\rm br}} \right)^{\Gamma_2-\Gamma_1} \right)^{-1}, 
\label{eq:g-sbr}
\end{equation}
where $E_0$ is 1~GeV. The photon indices $\Gamma_1$ below the break, $\Gamma_2$ above the break, a break energy ${E_{\rm br}}$, and a normalization factor $N_0$ are free parameters. 
The resulting TS are ${\rm TS_{Cutoff}} = -2\ln (L_{\rm PL}/L_{\rm Cutoff}) = 2.4$ and ${\rm TS_{BPL}} = -2\ln (L_{\rm PL}/L_{\rm BPL}) = 6.8$,
corresponding to a significance of $1.5\sigma$ and $2.1\sigma$ respectively.
There is an indication of the spectral steepening above a few GeV, but the possibility of a simple power law spectrum cannot be conclusively rejected.
The parameters obtained with the broken power law model are photon indices $\Gamma_1=1.4 \pm 0.5$, $\Gamma_2=2.5\pm0.15$, and $E_{\rm break}=1.0\pm 0.8$ GeV, with an integrated flux in 0.2--200 GeV of $(3.8\pm0.6)\times10^{-8}$~photon~cm$^{-2}$~s$^{-1}$. The photon index obtained with the simple power law is $2.14\pm0.05$.
The gamma-ray luminosity in 0.2--200 GeV is calculated as $1.3\times10^{34}~(d/1.3$~kpc)$^2$~erg~s$^{-1}$ using the result of the broken power law. 
The best-fit functions are represented in Figure~\ref{fig: sed}.

\subsection{Upper Limit on PSR~J0538$+$2817}

As shown in Figure~\ref{fig: smallFoV}, no gamma-ray counterpart is found at the position of PSR~J0538$+$2817 located near the center of the SNR.
An upper limit to the GeV flux from the pulsar is determined by performing a maximum likelihood fit in 0.2--200~GeV.
The model is the same model as in Section~\ref{sec: sed}, where S147 is modeled as a smoothly broken power-law function with the normalization, the energy break, and the photon indices below and above the break set free, except for adding a point source at a position of PSR~J0538$+$2817, specifically, ($\alpha$, $\delta$)=($84\fdg 604$, $28\fdg 286$)~\citep{S147_psr_R}.
The spectral shape of the pulsar is assumed to be an exponentially cut-off power-law function (see Equation~\ref{eq:g-cutoff}) with a photon index of 1.3 and a cutoff energy of 1.9~GeV~\citep{subL-PSR}.
Note that these assumed parameters are within typical values of the observed gamma-ray pulsars~\citep{FermiP-ctlg}.
The calculated upper limit on the flux (0.2--200~GeV) is $1.9\times10^{-9}\ {\rm photons}\ {\rm cm}^{-2}\ {\rm s}^{-1}$ at 90\% confidence level, 
corresponding to a luminosity limit of $4.7 \times 10^{32}$~($d/1.3$~kpc)$^2$~erg~s$^{-1}$.

\section{Discussion}
\label{sec: discuss}

\subsection{Modeling of Multiwavelength Spectra}
\label{sec: mulwave}

We found extended GeV gamma-ray emission spatially coincident with SNR S147 using 31 months of data acquired with the \emph{Fermi} LAT. 
The size of the gamma-ray emitting region is consistent with the shell of H$\alpha$ and radio emission ($\sim$200$\arcmin$).
Moreover, the gamma-ray emission exhibits a possible correlation with H$\alpha$ filaments and therefore also with synchrotron radio filaments given a tight correlation between 
the H$\alpha$ and radio maps \citep{S147_R1}. 
We did not find a gamma-ray source at the position of PSR~J0538$+$2817, which is considered to be the stellar remnant of the supernova explosion.

Let us consider the gamma-ray emissions from radio-emitting regions, 
which can be decomposed into filaments and diffuse regions \citep{S147_R1}. 
The diffuse and filamentary components explain the radio data below and above a break around 1.5~GHz ($\sim10^{-4}$~eV), respectively.
The diffuse radio emission is supposed to come from the shocked gas behind the blast wave 
which is propagating in intercloud medium, while the filaments 
are formed through radiative shocks driven in atomic clouds.
We calculate the gamma-ray spectrum of each zone expected from the radio spectra under some reasonable assumptions.
These components are separated only in spectral space at gamma-ray energies, since it is difficult to spatially resolve these components with the PSF of the LAT.
The distance to the remnant is assumed to be $d=1.3$ kpc, 
 which is the most likely distance to the pulsar PSR J0538$+$2817 that is 
supposed to be associated with S147 \citep{S147_psr_parallax,Chatterjee09,S147_psr_pm}.
Following this assumption, the radius of the remnant is 38~pc.
We also adopt a remnant age of $t_0=3 \times 10^4$~yr from the kinematic age of PSR~J0538$+$2817.
The distance, radius, and age have uncertainties of 20--30\% (as described in Section~1). 

After supernova explosions, 
when the swept-up gas becomes comparable to the ejecta mass and the blast wave slows, SNRs enter the Sedov-Taylor (adiabatic) phase~\citep[e.g.,][]{ST}.
The total energy of SNRs remains almost constant during the Sedov-Taylor phase,
because thermal and synchrotron radiation energy losses are negligible compared to the internal energy.
The radiative phase begins when radiative cooling dominates the energy loss of SNRs and the adiabatic approximation breaks down.
The transition time and radius to the radiative phase are 
\begin{eqnarray}
\label{eq: tr}
&&t_{\rm tr} = 1.3\times10^4 E_{51}^{3/14}n_{\rm ICM}^{-4/7}~~{\rm yr},\\
&&R_{\rm tr} = 14 \times E_{51}^{2/7}n_{\rm ICM}^{-3/7}~~{\rm pc}, 
\end{eqnarray}
where $E_{51}$ is the explosion kinetic energy in units of $10^{51}\ {\rm erg}$ 
and $n_{\rm ICM}$ is the density of the intercloud gas in units of ${\rm cm^{-3}}$~\citep{Cioffi1988}. 
If S147 is in the Sedov phase, the density of the intercloud gas should be $n_{\rm ICM} < 0.24E_{51}^{3/8}t_{\rm 30kyr}^{-7/4}$~cm$^{-3}$ ($\equiv n_{\rm cr}$) derived from the condition of $t_0 < t_{\rm tr}$, where $t_{\rm 30kyr} = t_0/(3\times10^4~{\rm yr})$.
This result is consistent with the density of the intercloud gas calculated by the relation in the Sedov-Taylor stage:
\begin{equation}
\label{eq: icm}
n_{\rm ICM} = 3.3\times10^{-2} E_{51} t_{\rm 30kyr}^2 R_{\rm 38pc}^{-5}~~{\rm cm}^{-3},
\end{equation}
where $R_{\rm 38pc}=R/(38~{\rm pc})$.
On the other hand, if we assume that S147 is in the radiative phase, the density of the intercloud gas should be $n_{\rm ICM} > n_{\rm cr}$.
This leads to a smaller remnant, incompatible with our preferred distance of 1.3~kpc.
In this paper, we assume S147 is in the Sedov phase.

The temporal evolution of the particle (protons/electrons) momentum distribution in diffuse and filamentary zones can be described by:
\begin{equation}
\label{eq: injection}
\frac{\partial N^{\rm d,f}_{e,p}}{\partial t}  = \frac{\partial}{\partial p} ( b_{e,p} N^{\rm d,f}_{e,p})  + Q^{\rm d,f}_{e,p},
\end{equation}
where $b_{e,p}=-dp/dt$ is the momentum loss rate, and $Q^{\rm d,f}_{e,p}(p)$ is the particle injection rate.
The super scripts ``d" and ``f" of the parameters represent the diffuse and the filamentary regions, respectively. 
To reproduce the spectral steepening in the radio band, 
we adopt phenomenological forms for injection distributions of protons and electrons:
\begin{eqnarray}
\label{eq: ep_dist}
Q^{\rm d,f}_{e,p}(p) = a^{\rm d,f}_{e,p} \left(\frac{p}{p_0}\right)^{-s} \exp{\left(-\left(\frac{p}{p_{\rm cut}^{\rm d,f}}\right)^2\right)},
\end{eqnarray}
where $p_0$ is 1~GeV~$c^{-1}$ and $p^{\rm d,f}_{\rm cut}$ is a cutoff momentum. Here we assume the injection rate is time independent for simplicity. 
To obtain the radiation spectra from the remnant, $N^{\rm d,f}_{e,p}(p, t)$ is numerically calculated using parameters in Table~\ref{tbl: basic}. 
We use $N^{\rm d}(p,t_0)$ and  $N^{\rm f}(p,t_0/2)$ 
to calculate the radiation spectra from the diffuse region and filaments, 
respectively.
The momentum losses for electrons include synchrotron radiation, bremsstrahlung, inverse compton (IC) scattering, and coulomb collisions, 
while those for protons include pion production losses, and coulomb collisions~\citep{Sturner97}.
In the case of the diffuse region, the adiabatic loss is also taken into account 
using the Sedov-Taylor evolution.
The gamma-ray emission mechanisms include the $\pi^0$-decay gamma rays due to high-energy protons, and bremsstrahlung and IC scattering processes by high-energy electrons. 
Calculations of the gamma-ray emission are done using the method described in \citet{FermiW51C}.
The interstellar field used for calculations of IC includes an infrared blackbody component ($kT_{\rm IR} = 3\times 10^{-3}$ eV, $U_{\rm IR} = 1\ \rm eV\ cm^{-3}$), an optical blackbody component ($kT_{\rm opt} = 0.25$ eV, $U_{\rm opt} = 1\ \rm eV\ cm^{-3}$), and the cosmic microwave background.
The infrared and optical components are set to reproduce the interstellar radiation field in the GALPROP code~\citep{Porter}.

\begin{deluxetable}{lrrlr}
\tabletypesize{\footnotesize}
\tablecaption{Basic Parameters of Multiwavelength Models \label{tbl: basic}}
\renewcommand{\arraystretch}{1.}
\tablewidth{0pt}
\tablehead{
\multicolumn{1}{l}{SNR dynamics}
}
\startdata
assumed parameters\\
~~Distance: $d$ & 1.3~kpc \\
~~Age: $t_0$  & $3 \times 10^4$~yr \\
~~Explosion energy: $E_{\rm tot}$ & (1--3) $\times 10^{51}$~erg\\[5pt]
dependent parameters\\
~~Radius: $R$    & 38~pc \\
~~Blast wave velocity: $v_b$ & 500~km~s$^{-1}$\\
\hline\\
\multicolumn{1}{l}{Filament gas properties}\\
\hline
assumed parameters\\
~~Preshock magnetic field: $B_{\rm atc}$ &3--10~$\mu$G \\
~~Gas density:      $n_{\rm Hf}$ &100--500~cm$^{-3}$ \\
~~Temperature:    $T_{\rm f}$ & $2\times10^4$~K\\
~~Shock velocity: $v_s$         &   100~km~s$^{-1}$\\[5pt]
dependent parameters\\
~~Magnetic field: $B_{\rm f}$ & $B_{\rm f}(n_{\rm Hf}, T_{\rm f}, B_{\rm atc}, v_s)\tablenotemark{a}$ \\
~~Preshock gas density: $n_{\rm atc}$ &2--6~cm$^{-3}$ \\
\hline\\
\multicolumn{1}{l}{Diffuse gas properties}\\
\hline
assumed parameters\\
~~Preshock magnetic field: $B_{\rm ICM}$ &1--5~$\mu$G\\[5pt]
dependent parameters\\
~~Magnetic field:  $B_{\rm d}$ &3--16~$\mu$G\\
~~Preshock gas density: $n_{\rm ICM}$ &0.03--0.1~cm$^{-3}$  \\
~~Gas density: $n_{\rm Hd}$ &0.1--0.4~cm$^{-3}$  \\
\enddata
\tablenotetext{a}{see Section~\ref{sec: mulwave}.}
\end{deluxetable}

The physical parameters used for the model calculations are summarized in Table~\ref{tbl: basic}. 
The blast wave velocity is determined by $v_b = 0.4R/t_0$~\citep[e.g.,][]{ST}. 
The density of the intercloud gas $n_{\rm ICM}$ is determined using Equation~\ref{eq: icm}.
Then, the post-shock density in the diffuse region is set by $n_{\rm Hd}=4  n_{\rm ICM}$. 
The magnetic field in the intercloud region is varied within typical values of $1$--$5\ \mu{\rm G}$,  when predicting the gamma-ray spectra.
The compressed magnetic field in the diffuse region $B_{\rm d}$ is determined as
$B_{\rm d} = \sqrt{2/3}(n_{\rm Hd}/n_{\rm ICM})B_{\rm ICM}$.

We assume that the filaments are formed by a radiative shock wave ($50\ {\rm km\ s^{-1}} < v_s < 200\ {\rm km\ s^{-1}}$)  driven into atomic clouds, 
which are denser than the intercloud medium.
The post-shock gas is subject to radiative cooling, forming compressed gas (i.e., filaments).
The gas density $n_{\rm Hf}$ and temperature $T_{\rm f}$ of the filaments are estimated from optical lines~\citep{S147_k2,Fes85} (see Table~\ref{tbl: basic}). 
We adopt $v_s = 100\ {\rm km\ s^{-1}}$ according to the optical observations~\citep{S147_k1,S147_k2,S147_k3}. 
The density of the atomic clouds $n_{\rm atc}$ is calculated as
\begin{eqnarray}
\label{eq: atc}
n_{\rm atc} &=& n_{\rm ICM} \left( \frac{v_b}{v_s} \right)^2 F\\
	           &=& 0.83 F E_{51} R_{\rm 38pc}^{-3} {v_{s7}}^{-2}~~{\rm cm}^{-3},
\end{eqnarray}
where $v_{s7}=v_s/(100\ {\rm km\ s^{-1}})$ and $F \simeq 3.2 - 4.8(v_s/v_b) + 2.6(v_s/v_b)^2$~\citep{MC75}.
The magnetic field strength in the filaments, $B_{\rm f}$, can be estimated from pressure balance~\citep{HM79}.
The shock ram pressure $P_{\rm ram} = n_{\rm atc}\mu_{\rm H}{v_s}^2$ should be balanced by magnetic or thermal pressure, 
where $n_{\rm atc}$ is the gas density of the atomic clouds (preshocked gas of the filaments), and $\mu_{\rm H}$ is the mass per hydrogen nucleus.
When magnetic pressure supports the filaments, 
the magnetic field strength is set by 
$P_{\rm ram} = {B_{\rm f}}^2/8\pi$. 
Using the relation of $B_{\rm f} = \sqrt{2/3}(n_{\rm Hf}/n_{\rm atc})B_{\rm atc}$, 
\begin{eqnarray}
\label{eq: Bfm}
B_{\rm f} \simeq 17 \times {v_{s7}}^{2/3} {B_{\rm atc-6}}^{1/3} {n_{\rm Hf}}^{1/3} ~~\mu {\rm G},
\end{eqnarray}
where $B_{\rm atc-6}=B_{\rm atc}/(1\ \mu{\rm G})$. 
The magnetic field in the atomic clouds $B_{\rm atc}$ is set in a range of a few times higher than a typical intercloud magnetic field.
When the thermal pressure supports the filaments, $P_{\rm ram}$ should be equated with  the thermal pressure $x_t n_{\rm Hf}kT_{\rm f}$, 
where $x_t  = 2.3$ assuming the filament gas is fully ionized and $k$ is the Boltzmann constant.
In this case,
\begin{eqnarray}
\label{eq: Bft}
B_{\rm f}   \simeq 60 \times {v_{s7}}^2 {B_{\rm atc-6}} T_{\rm f4}^{-1}~~~\mu {\rm G},
\end{eqnarray}
where $T_{\rm f4}$ is $T_{\rm f}/(10^4\ {\rm K})$. The lower value between Equations~\ref{eq: Bfm} and \ref{eq: Bft} 
determines $B_{\rm f}$, which is found to be $B_f \sim$100--300~$\mu$G for the adopted parameters.

We adjust the injection spectrum of  electrons $Q^{\rm d,f}_{e}(p)$ (Equation~\ref{eq: ep_dist}) to reproduce the observed radio spectrum, 
which is arguably the synchrotron radiation of the relativistic electrons.
Free parameters of $Q^{\rm d,f}_{e}(p)$ are $a^{\rm d,f}_{e}$, $p^{\rm d,f}_{\rm cut}$, and $s$.
The index $s$ is set at 1.5 to explain the observed radio synchrotron emission.
We use the same indices for the filamentary and diffuse components, while the normlaization and the cutoff momentum are set independently.
To predict the gamma-ray spectrum, 
 we consider two cases where the number ratio of the relativistic electrons to protons $K_{ep} \equiv a_e/a_p$ are 0.01 and 1. 
A value of $K_{ep} = 0.01$ is similar to what is locally observed for CRs at GeV energies~\citep[e.g.,][]{CRobs}. 

\begin{figure}[h]
\includegraphics[width=3.5in]{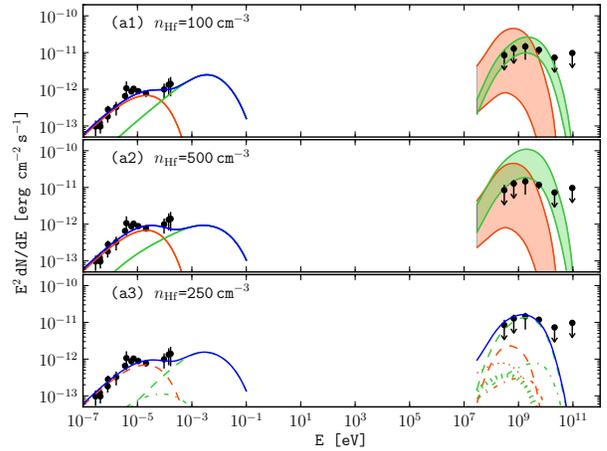}
\caption{
SEDs of S147 with model curves for three cases. Cases (a1), (a2), and (a3) represent $n_{\rm Hf}=100$, 500, and $250\ {\rm cm}^{-3}$, respectively (see Table~\ref{tbl:S147:model_pi0}).
$K_{ep}$ is assumed to be 0.01.  
 The black points represent observed data in the radio~\citep{S147_R1} and LAT bands.
The red and green lines represent diffuse and filamentary components, while the blue lines represent a sum of the two components.
The filled regions of cases (a1) and (a2) in the gamma-ray band show ranges of 
the gamma-ray spectra using parameters in Table~\ref{tbl:S147:model_pi0}.
Case (a3) shows sub-components of the models for a particular case. 
The synchrotron spectra from primary (dashed line) and secondary electrons (dot-dashed line) are drawn in the radio band, while the gamma-ray spectrum 
can be decomposed into 
 $\pi^0$-decay from the relativistic protons (dashed line), 
bremsstrahlung from primary (dot-dashed line) and secondary electrons (vertical-dashed line), and IC scattering from primary electrons (dotted line).
\label{fig:S147:mw_pi0}
}
\end{figure}

The corresponding gamma-ray spectra are predicted using physical parameters varied within a range described in Table~\ref{tbl: basic}. 
The gamma rays are dominated by the $\pi^0$-decay emission for $K_{ep} = 0.01$. 
Figure~\ref{fig:S147:mw_pi0} shows that the observed gamma-ray spectrum can be reproduced by using the parameters that are outlined above.
The color-filled regions in the gamma-ray band represent ranges of possible fluxes with physical parameters varied as described in Table~\ref{tbl:S147:model_pi0}. 
For case (a3), where the variable parameters are set at nearly central values in the expected ranges,
the total kinetic energies of protons ($W_p$) are calculated as $1.7 \times 10^{47}$~erg and $5.0 \times 10^{49}$~erg for the filaments and diffuse, respectively.
Although $W_p$ of the filaments is much smaller than that of the diffuse,
the observed gamma rays are dominated by the $\pi^0$-decay emission from the filaments due to the high density. 
The strong magnetic field of 210~$\mu$G also enhances radio synchrotron emission from the filaments.
Note that only the filament component can reproduce the observed gamma-ray spectral shape in any case.

\begin{figure}[h] 
\includegraphics[width=3.5in]{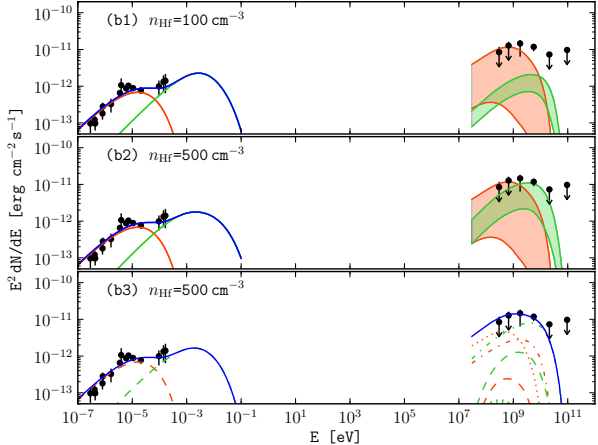}
\caption{\small 
Same as Figure~\ref{fig:S147:mw_pi0} but $K_{ep}$ is assumed as 1. 
Cases (b1), (b2), and (b3) represent $n_{\rm H}=100$, 500, and $500\ {\rm cm}^{-3}$, respectively (see Table~\ref{tbl:S147:model_br}).
\label{fig:S147:mw_brems}
}
\end{figure}

In the case of $K_{ep} =1$, the gamma-ray emission of the filaments is dominated by electron bremsstrahlung, while that of the diffuse region has significant contributions by both bremsstrahlung and IC scattering.
Figure~\ref{fig:S147:mw_brems} shows that the observed gamma-ray spectrum is under-predicted for $n_{\rm Hf} = 100\ {\rm cm}^{-3}$.
To reproduce the observed spectrum,  $n_{\rm Hf} \gtrsim 500\ {\rm cm}^{-3}$ is preferable. 
However, the total kinetic energy of the relativistic protons is only $W_p = 2.5 \times 10^{48}$~erg in case (b3).
This value is only one third of the energy content of Galactic CR protons in the volume of S147, $W_{\rm CR} = U_{\rm CR} V \simeq 7.5 \times 10^{48}$~erg, 
where $U_{\rm CR} \simeq 0.7$~eV~${\rm cm^{-3}}$ is the energy density of CR protons and $V = (4\pi/3) R^3$ is the volume of the SNR with a radius of $R=38$~pc.
This suggests that $K_{ep}$ is unlikely to be 1.

\begin{deluxetable*}{lcccccccc}
\tabletypesize{\scriptsize}
\tablecaption{Parameters of Multiwavelength Models ($K_{ep} = 0.01$) \label{tbl:S147:model_pi0}}
\tablewidth{0pt}
\tablehead{    
\colhead{} & \multicolumn{4}{c}{Parameters} &
& \multicolumn{3}{c}{Energetics}\\
\cline{2-5} \cline{7-9} \\
\colhead{Model} & \colhead{} & \colhead{${n}_{\rm H}$}
& \colhead{$B$} & \colhead{$p_{\rm cut}$}
& \colhead{}
& \colhead{$W_p$} 
& \colhead{$W_e$} & \colhead{$U_B$}\\
\colhead{} & \colhead{} & \colhead{[cm$^{-3}$]} 
& \colhead{[$\mu$G]} & \colhead{[GeV $c^{-1}$]} &
& \colhead{[$10^{48}$ erg]} & \colhead{[$10^{46}$ erg]} & \colhead{[eV cm$^{-3}$]} 
}
\startdata
(a1) filament
&  &  100          & 90--170 & 60        & &  0.29--0.78 & 0.29--0.84  & 200--720\\
~~~~~ diffuse
&  &  0.1--0.4    & 3--16     & 8--15   & &  22--350    &  42--550      &0.2--6.4\\[3pt]
(a2) filament
&  &  500          & 90--240 & 60        & &  0.11--0.63 & 0.080--0.56 & 200--1430\\
~~~~~ diffuse
& &  0.1--0.4 & 3--16        & 8--15    & &  22--350     &  42-550        &0.2--6.4\\[3pt]
(a3) filament
&  &  250          & 210 & 60 & & 0.17 & 0.15 & 1100\\
~~~~~ diffuse
&  &  0.2            & 10   & 10 & &  50  & 88 & 2.5\\
\enddata
\tablecomments{
The total kinetic energies ($E_{\rm kin}$) of radiating protons ($W_p$) and electrons ($W_e$) are calculated for $E_{\rm kin} > 50$~MeV and $E_{\rm kin} > 20$~MeV, respectively.}
\end{deluxetable*}
\begin{deluxetable*}{lcccccccc}
\tabletypesize{\scriptsize}
\tablecaption{Parameters of Multiwavelength Models ($K_{ep} = 1$)\label{tbl:S147:model_br}}
\tablewidth{0pt}
\tablehead{    
\colhead{} & \multicolumn{4}{c}{Parameters} &
& \multicolumn{3}{c}{Energetics}\\
\cline{2-5} \cline{7-9} \\
\colhead{Model} & \colhead{} & \colhead{${n}_{\rm H}$}
& \colhead{$B$} & \colhead{$p_{\rm cut}$}
& \colhead{}
& \colhead{$W_p$} 
& \colhead{$W_e$} & \colhead{$U_B$}\\
\colhead{} & \colhead{} & \colhead{[cm$^{-3}$]} 
& \colhead{[$\mu$G]} & \colhead{[GeV $c^{-1}$]} &
& \colhead{[$10^{46}$ erg]} & \colhead{[$10^{46}$ erg]} & \colhead{[eV cm$^{-3}$]} 
}
\startdata
(b1) filament
&  &  100          & 90--170 & 40 &    & 0.29--0.81 & 0.31--0.76 & 200--720\\
~~~~~ diffuse
&  &  0.1--0.4 & 3--16        & 8--15  &  & 24--390 &  46--610 &0.2--6.4\\[3pt]
(b2) filament
&  &  500          & 90--240 & 40 & & 0.22--0.81 & 0.18--0.76 & 200--1430\\
~~~~~ diffuse
& &  0.1--0.4 & 3--16        & 8--15  &  & 24--390    &  46--610 &0.2--6.4\\[3pt]
(b3) filament
&  &  500          & 90 & 40 & & 0.81 & 0.76 & 200\\
~~~~~ diffuse
&  &  0.4 & 4         & 14 &  & 250    &  390 & 0.4\\
\enddata
\tablecomments{
The total kinetic energies ($E_{\rm kin}$) of radiating protons ($W_p$) and electrons ($W_e$) are calculated for $E_{\rm kin} > 50$~MeV and $E_{\rm kin} > 20$~MeV, respectively.}
\end{deluxetable*}

\subsection{Reacceleration of Galactic Cosmic Rays as Sources of Gamma-ray Emitting Particles}
\label{sec: reacc}

The total energy of the relativistic electrons obtained for the most reasonable set of parameters (i.e., case (a3)), is $W_e \simeq 9 \times 10^{47}$~erg (see Table~\ref{tbl:S147:model_pi0}).
This value is only three times higher than the total energy of the Galactic CR electrons that was stored in the interstellar medium SNR S147 has swept up.
This indicates the possibility that the observed spectra can be  explained by the emission from only pre-existing CRs accelerated at the shock wave of the SNR shell.
In Section~\ref{sec: mulwave}, we modeled the radio and gamma-ray 
spectra without specifying the sources of energetic particles.
In this section, we calculate the emission from the relativistic particles produced by acceleration (i.e., reacceleration) of the pre-existing CRs, 
following the prescription given by \cite{Uchiyama10}.
The emission from two regions, filaments and diffuse, are considered in the same way as Section~\ref{sec: mulwave}.

The spectral distributions of the reaccelerated CRs are calculated as follows.
The number density of ambient CR protons and electrons, seeds of reacceleration, are adopted as the spectra of 
the observed Galactic CR protons $n_{{\rm GCR},p}(p)$ and electrons+positrons $n_{{\rm GCR},e}(p)$:
\begin{eqnarray}
&&n_{{\rm GCR},p}(p) = 4\pi J_p \beta^{1.5} p_0^{-2.76},\\
&&n_{{\rm GCR},e}(p) = 4\pi J_e \beta^{1.5} p_0^{-2}\ (1+p_0^2)^{-0.55},
\label{eq: GCRs}
\end{eqnarray}
where $p_0 = p/({\rm GeV}/c)$, $J_p = 1.9$~cm$^{-2}$~s$^{-1}$~sr$^{-1}$~GeV$^{-1}$, 
and $J_e = 2 \times 10^{-2}$~cm$^{-2}$~s$^{-1}$~sr$^{-1}$~GeV$^{-1}$ \citep{Sh07,GALPROP2}.
We assume that spectral distributions of the pre-existing relativistic particles in the atomic clouds (i.e., preshocked gas of the filaments) are also the same as those of the Galactic CRs.
At the shock wave, the CRs described above are assumed to be accelerated according to the theory of diffusive shock acceleration~\citep{DSA}.
The number density of the reaccelerated CRs as a function of momentum $n_{\rm acc}(p)$ is described as
\begin{equation}
n_{\rm DSA}(p) = (\alpha + 2)\ p^{-\alpha} \int_0^p dp^{\prime} n_{\rm GCR}(p^{\prime})\ p^{\prime(\alpha-1)},
\label{eq:dis:umdl:acc}
\end{equation}
where $n_{\rm GCR}$ is the number density of pre-shock CRs, 
$\alpha \equiv (r_{\rm sh}+2)/(r_{\rm sh}-1)$ and $r_{\rm sh}$ is the shock compression ratio.
Here we adopt $r_{\rm sh}$ to be 4, assuming the specific heat ratio of non-relativistic monoatomic gas ($=5/3$) and a strong shock (Mach number $\gtrsim 10$).
Given the finite acceleration time, the maximum attainable momentum of the accelerated particles is also finite.
Here we introduce a cutoff momentum ($p_{\rm c\_acc}$) for the spectra of the accelerated particles.
Then, the number density of particles accelerated at the shock wave is described as
\begin{eqnarray}
n_{\rm acc}(p) = n_{\rm DSA}(p) \times \exp{\left(-\left(\frac{p}{p_{\rm c\_acc}}\right)^2\right)}.
\label{eq:dis:umdl:DSA}
\end{eqnarray}
Since the diffuse region is behind the shock wave, the number density of relativistic particles in the diffuse region $n^{\rm d}_{e,p}(p)$ has the spectral distribution of Equation~\ref{eq:dis:umdl:DSA}. 
On the other hand, the filaments are formed as the shocked gas is compressed radiatively.
In this process, the particles frozen in the gas are heated and gain energy as $p \rightarrow s^{1/3}p$, 
where $s \equiv (n_{\rm Hf}/n_{\rm \bf atc})/r_{\rm sh}$, and the density increases by a factor of $s$~\citep{BC82}. 
Therefore, the number density of the accelerated and compressed particles in the filaments is calculated as $n^{\rm f}_{e,p}(p) = s^{2/3} n_{\rm acc}(s^{-1/3}p)$.

To calculate the multiwavelength spectrum, 
the same procedure and parameters as Section~\ref{sec: mulwave} are applied except for the injection rates of the relativistic particles:
\begin{eqnarray}
\label{eq: ep_dist2}
Q^{\rm d}_{e,p}(p) &=& (1-f)V (n_{\rm ICM}/n_{\rm Hd}) \times n^{\rm d}_{e,p}(p) /{t^{\rm d}_{\rm int}},\\ 
Q^{\rm f}_{e,p}(p) &=& f V (n_{\rm atc}/n_{\rm Hf}) \times n^{\rm f}_{e,p}(p) /{t^{\rm f}_{\rm int}},
\end{eqnarray}
where $t^{\rm d}_{\rm int} = t_0$ and $t^{\rm f}_{\rm int} = t_0/2$ are the total injection time.
The filling factor of the preshocked atomic gas is defined as $f \equiv V_{0}/V$, where $V_0$ and $V$ are volumes of the atomic clouds and the SNR, respectively.
In this scenario, the relativistic particle distributions are almost fixed unlike those in Section~\ref{sec: mulwave}.
Free parameters of the particle distributions are only $f$ and $p^{\rm d,f}_{\rm c\_acc}$. 
In the calculation of the diffuse components, the parameters $B_{\rm d}$ and $p^{\rm d}_{\rm c\_acc}$ are uniquely determined to fit the radio observational data.
The gas density $n_{\rm Hd}$, which affects the calculated flux of the gamma-ray emission, ranges within the values in Table~\ref{tbl: basic}.
We note that the gamma-ray emission in the diffuse region is negligible within this range of $n_{\rm Hd}$, compared with the emission in the filamentary region.
On the other hand, the gas properties of the filaments have free parameters of $n_{\rm Hf}$ and $n_{\rm atc}$.
When we set $n_{\rm Hf}$ and $n_{\rm atc}$ within the values in Table~\ref{tbl: basic}, 
the parameters $B_{\rm f}$, $p^{\rm f}_{\rm c\_acc}$, and $f$ are uniquely determined to fit the observational data in the gamma-ray and radio bands.

\begin{figure}[h] 
\includegraphics[width=3.5in]{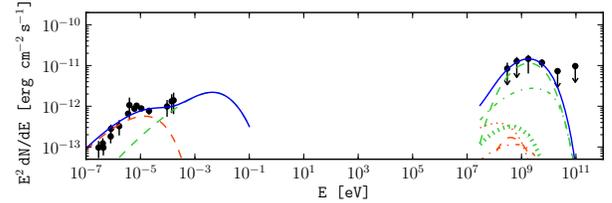}
\caption{\small 
SEDs of SNR~S147 with model curves of emission from the reaccelerated CRs in the case of $n_{\rm Hd} = 0.4\ {\rm cm}^{-3}$, $n_{\rm Hf} = 250\ {\rm cm}^{-3}$, and $n_{\rm atc} = 6\ {\rm cm}^{-3}$.
The distance to the SNR is assumed to be $d = 1.3$~kpc.
The observed data in the gamma-ray and the radio bands (black points) are the same as Figure~\ref{fig:S147:mw_pi0}.
The red and green lines represent diffuse and filamentary components, while the blue lines represent a sum of both fluxes.
The radio emission is modeled by synchrotron radiation from the relativistic electrons (dashed line).
The gamma-ray emission is explained with a combination of $\pi^0$-decay from the relativistic protons (dashed line), 
bremsstrahlung from primary (dot-dashed line) and secondary electrons (vertical-dashed line), and IC scattering from primary electrons (dotted line).
\label{fig:umdl:sed}
}
\end{figure}

Figure~\ref{fig:umdl:sed} shows the spectra in the case of $n_{\rm Hf} = 250\ {\rm cm}^{-3}$, $n_{\rm atc} = 6\ {\rm cm}^{-3}$, and $n_{\rm Hd} = 0.4\ {\rm cm}^{-3}$.
The cases of different values of the parameters are summarized in Table~\ref{tbl: model_reacc}.
Note that the spectra are almost the same for the values of the parameters in the table.
The gamma-ray flux is dominated by the $\pi^0$-decay emission from the dense filaments due to high densities of gas and CRs as in the case of Section~\ref{sec: mulwave}.
This is consistent with the result in Section~\ref{sec: detection}, which indicates the spatial correlation between the GeV gamma-ray and the filamentary H$\alpha$ emissions.
The obtained magnetic pressures are consistent with the values in Table~\ref{tbl: basic}, except for the case of $n_{\rm Hf} = 100\ {\rm cm}^{-3}$ where $B_{\rm f} = 70\ \mu{\rm G}$ is slightly smaller than that in the table.
In all cases, the filaments are supported by thermal pressure.
While the intensity of emission can be explained well,
the observed spectral index is harder than the calculated one at lower energy below the break in the radio band.
This discrepancy could be attributable to a cutoff around 100~MHz ($4\times10^{-7}\ {\rm eV}$) due to free-free absorption, which makes 
the spectral index harder than predicted by our model.

\begin{deluxetable*}{lcccccccccccc}
\tabletypesize{\scriptsize}
\tablecaption{Parameters for the Model of Reaccelerated Cosmic-ray Emission ($d=1.3$~kpc)\label{tbl: model_reacc}}
\tablewidth{0pt}
\tablehead{    
\colhead{}&\colhead{}&\multicolumn{5}{c}{Parameters} &
& \multicolumn{4}{c}{Energetics}\\
\cline{2-3} \cline{5-7} \cline{9-12} \\
\colhead{} & \colhead{${n}_{\rm H}$\tablenotemark{a}} & \colhead{${n}_{\rm atc}$\tablenotemark{a}} & \colhead{}
& \colhead{$B$} & \colhead{$p_{\rm c\_acc}$}
& \colhead{$f$}
& \colhead{}
& \colhead{$W_p$} 
& \colhead{$W_e$} & \colhead{$U_p$} & \colhead{$U_B$}\\
\colhead{} & \colhead{[cm$^{-3}$]} & \colhead{[cm$^{-3}$]} & \colhead{} 
& \colhead{[$\mu$G]} & \colhead{[GeV $c^{-1}$]} & \colhead{[\%]} &
& \colhead{[$10^{48}$ erg]} & \colhead{[$10^{46}$ erg]} & \colhead{[eV cm$^{-3}$]} & \colhead{[eV cm$^{-3}$]} 
}
\startdata
filament
&    100  &2--6   &&   70 & 40--60 & 0.6--0.8   && 0.36  &  1.8   &   80--270 & 120\\
&    250  &2--6   && 120 & 30--40 & 0.2--0.3   && 0.15  & 0.68  & 220--820 & 360\\
&    500  &2--6  &&  190 & 20--30 & 0.10--0.13   && 0.080 & 0.30 & 480--1600 & 900\\
diffuse
& 0.1--0.4&-- &&    13 &  8   & --    && 6.6    & 71 & 2.5 & 4\\
\enddata
\tablenotetext{a}{The densities of the gas are taken from Table~\ref{tbl: basic}. See Section~\ref{sec: reacc} for details.}
\tablecomments{
The total kinetic energies ($E_{\rm kin}$) of radiating protons ($W_p$) and electrons ($W_e$) are calculated for $E_{\rm kin} > 50$~MeV and $E_{\rm kin} > 20$~MeV, respectively.}
\end{deluxetable*}
\begin{deluxetable*}{lcccccccccccc}
\tabletypesize{\scriptsize}
\tablecaption{Parameters for the Model of Reaccelerated Cosmic-ray Emission ($d = 0.88~{\rm kpc}$)\label{tbl: model_reacc2}}
\tablewidth{0pt}
\tablehead{    
\colhead{}&\colhead{}&\multicolumn{5}{c}{Parameters} &
& \multicolumn{4}{c}{Energetics}\\
\cline{2-3} \cline{5-7} \cline{9-12} \\
\colhead{} & \colhead{${n}_{\rm H}$} & \colhead{${n}_{\rm atc}$} & \colhead{}
& \colhead{$B$} & \colhead{$p_{\rm c\_acc}$}
& \colhead{$f$}
& \colhead{}
& \colhead{$W_p$} 
& \colhead{$W_e$} & \colhead{$U_p$} & \colhead{$U_B$}\\
\colhead{} & \colhead{[cm$^{-3}$]} & \colhead{[cm$^{-3}$]} & \colhead{} 
& \colhead{[$\mu$G]} & \colhead{[GeV $c^{-1}$]} & \colhead{[\%]} &
& \colhead{[$10^{48}$ erg]} & \colhead{[$10^{46}$ erg]} & \colhead{[eV cm$^{-3}$]} & \colhead{[eV cm$^{-3}$]} 
}
\startdata
filament
&    250  & 10   && 120 & 50 & 0.6   && 0.070  & 0.31  & 100 & 360\\
~~dependence on $d$ & -- & $d^{-2.5}$ && -- & $d^{-0.8}$ & $d^{-1.7}$ && $d^2$ & $d^2$ & $d^{3.2}$ & --\\ \\
diffuse
& 1.8&-- &&    17 &  7   & --    && 1.8    & 21 & 2.2 & 7\\
~~dependence on $d$ & $d^{-5}$ & -- && $d^{-0.7}$ & $d^{0.35}$ & -- && $d^3$ & $d^3$ & --  & $d^{-1.4}$\\
\enddata
\tablecomments{
The total kinetic energies ($E_{\rm kin}$) of radiating protons ($W_p$) and electrons ($W_e$) are calculated for $E_{\rm kin} > 50$~MeV and $E_{\rm kin} > 20$~MeV, respectively.}
\end{deluxetable*}

In this paper, we assume the compression ratio of 4 based on the specific heat ratio of non-relativistic gas ($=5/3$).
The compression ratio is expected to increase if we consider the effect of the relativistic gas whose specific heat ratio is $4/3$.
The precise treatment of this effect is beyond the scope of this paper.
Here, let us consider the case of a compression ratio of 7 based on the specific heat ratio of the relativistic gas, i.e., CRs.
In this case, the calculated spectral index of the particle energy distribution becomes harder by $\sim0.5$.
The reacceleration model can still explain the observed multiwavelength spectra when we set the values of $f$ and $B_{\rm d}$ to $\sim60$\% of those in the default case where the compression ratio is 4. 
The harder spectra explain the observed radio data better than the default case.

Middle-aged SNRs interacting with molecular clouds 
constitute the dominant class 
of SNRs detected by the \emph{Fermi} LAT \citep[see][]{Uchiyama11}. 
It has been proposed by \citet{Uchiyama10} that the radiative filaments formed through interactions between molecular clouds and 
the SNR blast wave can account for the high gamma-ray luminosity of these 
SNRs (the Crushed Cloud model).
 In most cases, the reacceleration of Galactic CRs is sufficient to 
supply the required CR density in the filaments. 
The scenario discussed in this section can be regarded as the atomic cloud 
version of the Crushed Cloud model, which is indeed 
 insensitive to the physical parameters of a pre-shock cloud.
The GeV gamma-ray emission from the Cygnus Loop  \citep{FermiCygLoop} 
may be explained by the Crushed Cloud model as well. 

\subsection{Dependence on the Distance}

In Section~\ref{sec: reacc}, the distance to S147 is adopted as that to the plausibly associated pulsar.
However, the absorption lines of the B1e star indicate a different distance.
In this section, we consider a dependence on the distance for the reacceleration model as described in Section~\ref{sec: reacc}.

We first evaluate the dependence on the distance of the parameters in the diffuse region.
The gas density in the diffuse region is estimated to be $n_{\rm Hd} = 4n_{\rm ICM} \propto d^{-5}$ using Equation~\ref{eq: icm}.
The total energy of the accelerated particles is $W_{e,p}^{\rm d} \propto \sim n^{\rm d}_{e,p}(p)V \propto d^3$.
The energy density $U^{\rm d}_p = W^{\rm d}_p/V$ is thus independent of the distance.
The emissivity of the synchrotron emission is proportional to $B^{(m+1)/2}$, assuming that the radiating electrons obey a power-law distribution with the index of $m$.
Here we approximate the distributions of the accelerated particles by a power-law distribution with an index of 1.8.
The flux of the calculated synchrotron emission should explain the radio observational data for any distance; ${\rm (the\ calculated\ flux)} \propto B_{\rm d}^{(m+1)/2} W^{\rm d}_e/d^2 = {\rm constant}$.
The magnetic field $B_{\rm d}$ is thus proportional to $\sim d^{-0.7}$. The photon energy of the synchrotron emission is approximately proportional to $\gamma^2B$.
To explain the observational break in the radio band, $p^{\rm d}_{\rm c\_acc} \propto \sim\gamma \propto B_{\rm d}^{-0.5} \propto d^{0.35}$.
Next, the parameters of the filaments are considered.
The filamentary gas density $n_{\rm Hf}$ is derived from the optical observation, and independent of the distance.
On the other hand, the gas density of the atomic cloud is $n_{\rm atc} \propto d^{-2.5}$ using Equation~\ref{eq: atc}, 
where we use an approximate proportionality of $F \propto d^{0.5}$ around $d=1.3\ {\rm kpc}$.
The observed gamma-ray emission is reproduced by the $\pi^0$-decay emission in the filaments; ${\rm (the\ calculated\ flux)} \propto n_{\rm Hf}W^{\rm f}_p/d^2 = {\rm constant}$,
which leads the relation $W^{\rm f}_p \propto d^2$.
Since $W^{\rm f}_e/d^2 \propto W^{\rm f}_p/d^2$ is independent of the distance, 
the magnetic field $B_{\rm f}$ should also be independent of the distance to reproduce the observed synchrotron emission in the radio band.
Here we define the compression ratio $r_{\rm comp}$ as  $n_{\rm Hf}/n_{\rm atc} \propto d^{2.5}$.
The number distribution of the accelerated particles has a dependence on distance of 
$n^{\rm f}_{e,p}(p) = s^{2/3}n_{\rm acc}(s^{-1/3}p) \propto \sim s^{(2+m)/3} \propto d^{3.2}$, where $s \equiv r_{\rm comp}/r_{\rm sh} \propto d^{2.5}$.
The energy density is thus estimated to be $U^{\rm f}_p \propto n^{\rm f}_{e,p}(p) \propto d^{3.2}$.
The total energy is described as $W^{\rm f} \propto fV/r_{\rm comp} \times n^{\rm f}_{e,p}(p)$, which yields the filling factor of $f \propto d^{-1.7}$.
The cutoff momentum $p^{\rm f}_{\rm c\_acc}$ is determined by the observed gamma-ray data.
Since $p^{\rm f}_{\rm c\_acc}$ is approximately proportional to the gamma-ray cutoff calculated by the $\pi^0$-decay emission, 
the cutoff has a dependence on distance of $p^{\rm f}_{\rm c\_acc} \propto s^{-1/3} \propto \sim d^{-0.8}$.

Table~\ref{tbl: model_reacc2} shows one of the parameter sets of the reacceleration model for the case of $d=0.88~{\rm kpc}$.
The dependence on the distance $d$ is also summarized in it.
The observed intensity of the gamma-ray and radio emission is also reproduced by the reacceleration model for $d=0.88~{\rm kpc}$.
We note that the contribution of gamma rays in the diffuse region is not negligible below a few~GeV when $n_{\rm Hd}$ is greater than $\sim 2\ {\rm cm^{-3}}$.
In any case, the gamma-ray emission of the filaments is needed to explain the observed data above a few~GeV.

\section{Conclusions}

We have presented results for the GeV gamma-ray observations of the region around SNR S147 using about 31 months of data accumulated by the \emph{Fermi} LAT.
The gamma-ray emission is spatially extended, being consistent with the size of S147 ($\sim200\arcmin$).
There is no indication that the gamma-ray emission comes from the associated pulsar PSR J0538$+$2817.
The best fit to the LAT data is obtained using the H$\alpha$ image as a SNR spatial template rather than simple geometrical shapes.
Comparisons between gamma-ray and H$\alpha$ fluxes indicate a possible correlation between them, 
suggesting that the gamma rays come from the thin filaments observed in the H$\alpha$ and radio bands.
The observed energy spectrum between 0.2--200~GeV has a indication of spectral steepening; 
a smoothly broken power law provides a better fit than a simple power law at $2\sigma$ significance, which means the possibility of a simple power law cannot be rejected. 
The gamma-ray luminosity amounts to $1.3 \times 10^{34}~(d/1.3$~kpc)$^2$~erg~s$^{-1}$.

The LAT spectrum is best explained by $\pi^0$-decay gamma rays from relativistic protons in the dense filaments. 
We find that the reacceleration of pre-existing CRs and subsequent adiabatic compression in the filaments is sufficient to provide 
the required energy density of high-energy electrons and protons.
There are two main distance estimates to S147: from the pulsar association (1.3~kpc) or from the absorption lines ($<0.88$~kpc). 
We consider the cases of 1.3~kpc and 0.88~kpc, and the gamma-ray emission can be explained in the same framework for either distance.
SNR S147 offers a firm example of the realization of the Crushed Cloud model, and supports the importance of dense filaments in SNRs as gamma-ray production sites.

\acknowledgments
We thank the anonymous referee for her/his suggestions which have improved this paper.
The \textit{Fermi} LAT Collaboration acknowledges generous ongoing support
from a number of agencies and institutes that have supported both the
development and the operation of the LAT as well as scientific data analysis.
These include the National Aeronautics and Space Administration and the
Department of Energy in the United States, the Commissariat \`a l'Energie Atomique
and the Centre National de la Recherche Scientifique / Institut National de Physique
Nucl\'eaire et de Physique des Particules in France, the Agenzia Spaziale Italiana
and the Istituto Nazionale di Fisica Nucleare in Italy, the Ministry of Education,
Culture, Sports, Science and Technology (MEXT), High Energy Accelerator Research
Organization (KEK) and Japan Aerospace Exploration Agency (JAXA) in Japan, and
the K.~A.~Wallenberg Foundation, the Swedish Research Council and the
Swedish National Space Board in Sweden.

Additional support for science analysis during the operations phase is gratefully
acknowledged from the Istituto Nazionale di Astrofisica in Italy and the Centre National d'\'Etudes Spatiales in France.

\bibliographystyle{apj}
\bibliography{bibs_S147}

\end{document}